\documentclass[aps, showpacs, prb,twocolumn,
a4paper, notitlepage]{revtex4-1}
\usepackage{graphicx}
\usepackage{amsmath, amsfonts, amssymb, bm}
\usepackage{color} 

\usepackage{amsmath,array,amssymb,pst-all}
\usepackage{dsfont} 
\usepackage{ bbold } 

\newcommand{\bra}[1]{\left\langle #1 \right|}
\newcommand{\ket}[1]{\left| #1 \right\rangle}
\newcommand{\opj}{\hat{\cal J}}

\newcommand{\sdp}{\sin{\delta}}

\newcommand{\cde}{\cos{\frac{\delta}{2}}}
\newcommand{\sde}{\sin{\frac{\delta}{2}}}

\newcommand{\sfp}{\sin{(\phi_A+\phi_B)}}

\newcommand{\ctsp}[1]{\cos{\theta_{#1}}}
\newcommand{\stsp}[1]{\sin{\theta_{#1}}}
\definecolor{armygreen}{rgb}{0, 0.53, 0.13}

\begin{document}

\title{Quantum Locality in Game Strategy}

\author{Carlos A. Melo-Luna$^{1,2,3}$,  Cristian E. Susa$^{1,2,4}$, Andr\'es F. Ducuara$^{1,2}$, Astrid Barreiro$^{2}$ 
	\& John H. Reina$^{1,2}$}
\email{john.reina@correounivalle.edu.co} 
\address{$^{1}$Centre for Bioinformatics and Photonics---CIBioFi, Calle 13 No.~100-00, Edificio 320 No.~1069, 760032 Cali, Colombia}
\address{$^{2}$Departamento de F\'isica, Universidad del Valle, 760032 Cali, Colombia}
\address{$^{3}$
Experimental Physics IV, University of Bayreuth, 95440 Bayreuth, Germany
}
\address{$^4$Clarendon Laboratory, Department of Physics, University of Oxford, Oxford OX1 3PU, UK}

\date{\today}

\begin{abstract}
\textbf{Game theory is a well established branch of
mathematics whose formalism has a vast range of applications from the social sciences, biology,  to economics.  Motivated by  quantum information science,
there has been a leap in the formulation of novel game strategies that lead to new (quantum Nash) equilibrium points whereby  players in some classical games are always outperformed if  sharing and processing joint information ruled by the laws of quantum physics is allowed.
We show that, for  a bipartite non zero-sum game,  input {\it local} 
quantum correlations, and  {\it separable} states in particular,
suffice to achieve an advantage over any strategy that uses classical resources, thus dispensing with quantum nonlocality, entanglement, or even discord between the players' input states. This highlights the remarkable key role played by  pure quantum coherence at powering  some protocols. 
Finally, we propose an experiment that uses separable states and basic photon interferometry  to  demonstrate the  locally-correlated quantum advantage.}
\end{abstract}
\maketitle

In 1944, von Neumann developed a formal framework of game theory~\cite{von44}, namely of understanding the dynamics of competition and cooperation between two or more competing parties that hold particular interests.  In another seminal work,  twenty years later, Bell discovered the intrinsic, fundamental  nonlocal character of quantum theory~\cite{bell},  the fact that there exist quantumly correlated (entangled) particles whose measurement  gives results that are impossible in classical physics---the so-called violation of Bell inequalities~\cite{chsh,alain}.
Such Bell nonlocality and entanglement turned out to be of key relevance in the development of quantum information science and technology~\cite{NiCh10}. In fact,
quantisation protocols for strategy games exemplify a physical process whereby entanglement or nonlocality are used as a fundamental resource~\cite{meyer,jens,pappa,brunner,iqbal,Zh11,PSWZ07,Zu12,simon2,schmid,MaWe00,Du02,LJ03,FA03}. This establishes a connection between game theory and quantum information  and, as such, introduces the existence of certain advantages over the foregoing classical results~\cite{meyer,jens,pappa,brunner,iqbal,Zh11,PSWZ07,Zu12,simon2,schmid,MaWe00,Du02,LJ03,FA03}, and extends the set of cases that find solution to the interaction formalism~\cite{von44,Me91,osborne02} into the quantum realm~\cite{meyer,jens}.
Such quantum features are reflected, e.g.,  in the increase of efficiency and payoffs,  emergence of new equilibria, and novel  game strategies  which are simply not possible in the classical domain~\cite{LJ03, FA03, Du02, PSWZ07}. 
These achievements signalled an interest about the nature of  such a quantum advantage, and introduced questions related to the properties of physical systems and the mathematical structure that underlies the novel game strategies~\cite{pappa,brunner,iqbal,Zh11, PSWZ07,Zu12}. Advantages of different kind became evident  when  quantisation rules were applied to different sort  of games, and most of these scenarios  pointed out quantum  entanglement as a  precursor of such effects~\cite{meyer,jens,simon2,schmid,MaWe00,Du02,LJ03,FA03}.
 
\begin{figure}[ht!]
\vspace{-0.5cm}
        \begin{center}
         \includegraphics[scale=0.21]{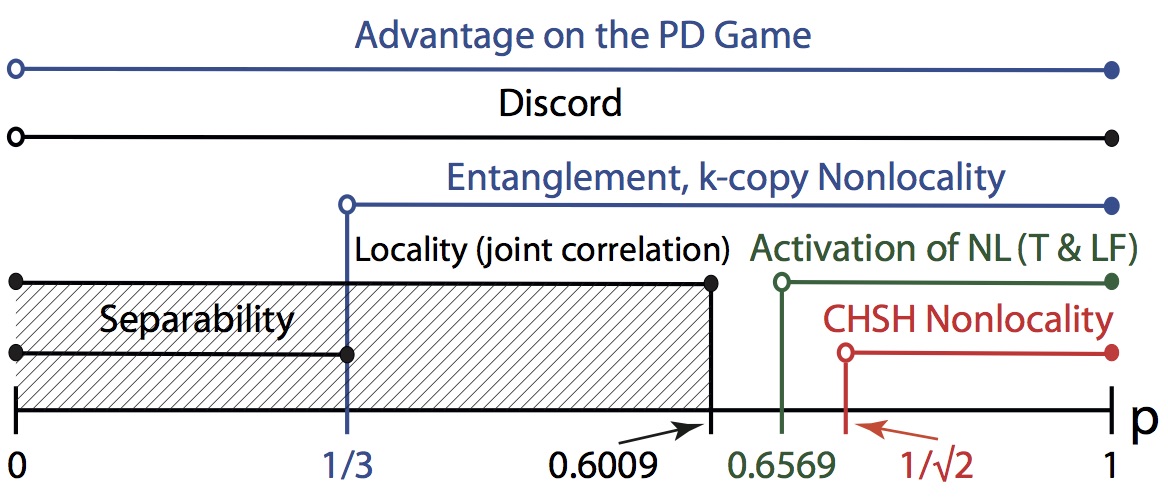}
        \caption{\label{properties}{\bf Some quantum properties for  two-qubit Werner-like states.}~The schematics highlights locality (for the joint correlation), entanglement, CHSH-nonlocality, $k$-copy nonlocality, 
        activation of nonlocality through tensoring and local filtering, and discord, for the Werner-like states 
         $\rho_{W{\text -}l}(p)=p| \psi \left>\right<\psi|+\frac{(1-p)}{4} \mathds{1}\otimes \mathds{1}$,  $\left| \psi \right>=\frac{1}{ \sqrt 2}\left(\left| 00 \right>+i\left| 11 \right>\right)$, $0 \leq p \leq 1$.  These states can lead to a PD game  advantage  in the whole $p$-region.}
\end{center}
\end{figure}

Furthermore, Bell nonlocality has been recently shown to provide an advantage when deciding conflicting interest games~\cite{pappa, brunner, iqbal}.
In this regard, and mostly inspired by strategies of this sort, the activation of quantum nonlocality has been put forward~\cite{NL2014, Ducu2016}. In particular, $k$-copy nonlocality or superactivation~\cite{palazuelos}, and activation of nonlocality through tensoring and local filtering~\cite{liang12}, although seminal for protocols based on nonlocality (e.g., quantum cryptography), are ultimately limited by the presence of entanglement~\cite{NL2014}. This said, 
here we explore other kind of correlations
that highlight local states  as a possible resource for introducing a quantum advantage (see Fig.~\ref{properties}, shaded region).
In particular, we ask whether there is, beyond entanglement or nonlocality, another underlying fundamental quantum feature as quantum coherence that warrants the emergence of the above-mentioned  advantages. This consideration is also motivated by a recent experimental demonstration of a {\it zero-sum} game that exhibits a quantum gain for players that do not share entanglement~\cite{Zu12}. 

The Prisoners' Dilemma  (PD) game is a celebrated 
bipartite non-zero sum
game in classical game theory~\cite{Me91,osborne02} whereby 
 two parties, say Alice (\emph{A}) and Bob (\emph{B}), have to decide between two strategies in an  independent way:  to defect (\emph{D}) or cooperate (\emph{C}). The retribution to the players decision is conditioned to the pair of choices, as shown in TABLE~\ref{tab:PayOff}. 
The classical PD game reveals the existence of a set of strategies from which unilateral movement of the players 
diminishes their payoff---a Nash equilibrium (NE)--- and  a set of strategies whereby the players simultaneously do best---a Pareto optimal~\cite{Me91}. The dilemma arises due to the choice problem between the equilibrium and the optimal gain.
\begin{table}[h]
\caption[caption]{Payoff matrix for the PD game. The
first (second) entry in the parenthesis denotes  Alice's (Bob's) payoff. In the classical game, the strategy (\emph{C},\emph{C}) defines a Pareto optimal (joint maximum gain), and (\emph{D},\emph{D}) a Nash equilibrium.}
\begin{tabular*}{8.7cm}{c @{\extracolsep{\fill}}c@{\extracolsep{\fill}}c}
	\hline
	\hline
        Alice\textbackslash Bob&$C$ &  $D$
        \\ \hline
        $C$& (3,3) &  (0,5)\\
        $D$&   (5,0) & (1,1)\\
	\hline
	\hline
\end{tabular*}
\label{tab:PayOff}
\end{table}

\begin{figure}[t]
        \begin{center}
         \includegraphics[scale=0.16]{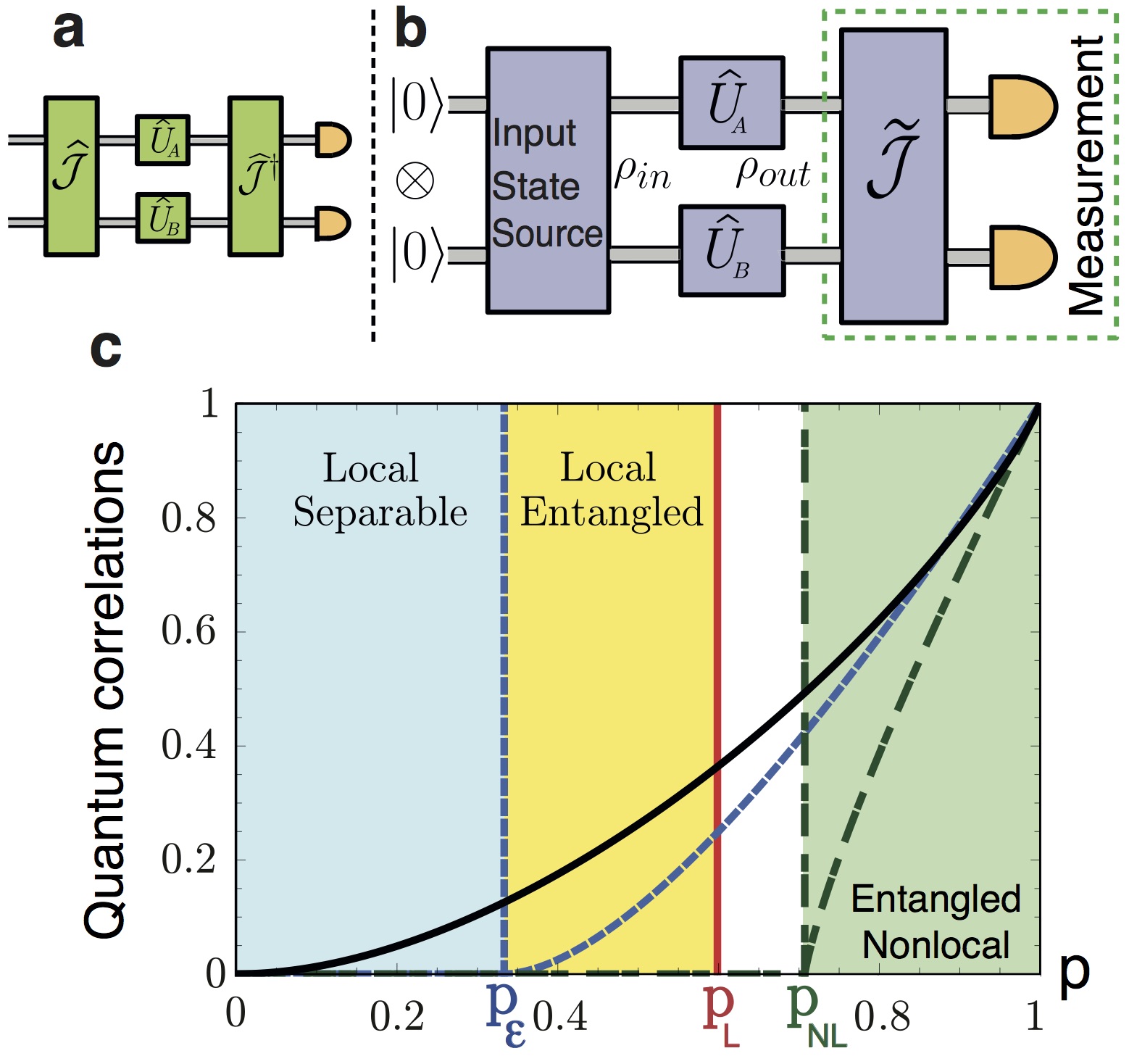}
        \caption{\label{protocol} {\bf Quantum Prisoners' Dilemma setup and classification of input correlations.} ({\bf a}) Eisert {\it et al.} two players game protocol~\cite{jens}, ({\bf b}) our setup uses a source of input $\rho_{in}\equiv\rho_{in}(p,\delta)$ (e.g., Werner-like) states, one qubit gates to represent the players' moves, and the measuring process (dashed rectangle). The measurement is taken as the projection onto the basis generated by $\tilde{\cal{J}}\equiv\tilde{\cal{J}}(\delta)$ in the usual 4-dimensional basis, 
        ({\bf c})  quantum correlations of input  $\rho_{W{\text -}l}(p)$ states: discord $ \mathcal{D}$ (solid-black), entanglement of formation $ \mathcal{E}$ (dashed-blue), and CHSH-nonlocality (doubly-dashed green).}
        \end{center}
        \vspace{-0.5cm}
\end{figure}

The PD sum game has been extended to the  quantum domain by Eisert {\it et al.}~\cite{jens}, who proposed the use of initial maximally entangled states and  unitary operators to define a strategy  of purely quantum character  that removes the decision dilemma~\cite{jens}. Thus, the interaction between players can be cast in a quantum circuit that generates, via the action of a two-qubit operator $\opj(\delta)$, an initial state of the form: 
\begin{equation}
\ket{\psi_{in}(\delta)}=\cde\ket{00}+i\sde\ket{{11}},
\label{EWLstate}
\end{equation}
where $\delta \in [0,\pi/2]$.  Here, the possible outcomes of the classical strategies \emph{C} and \emph{D} are assigned to the computational basis states $\ket{0}$ and $\ket{1}$, respectively, and the strategy space of each player has a Hilbert space structure that couples through a tensor product. 
In Fig.~\ref{protocol}(a), the operator $\opj(\gamma)=\exp\{i\gamma\hat{D}\otimes\hat{D}/2\}$ generates input entangled states\cite{jens}. 
We introduce the operator $\tilde{\cal{J}}(\delta)$, such that  $\tilde{\cal{J}}= \opj^\dagger$, as sketched in Fig.~\ref{protocol}(b).
After that, the players execute, unilaterally, their movements acting with the  unitary parameterised operator ($i=A, B$),
\begin{equation}
\hat{U}_{i}(\theta_i,\phi_i)=
        \left(
        \begin{array}{cc}
        e^{i\phi_i}\cos{\frac{\theta_i}{2}}&
        \sin{\frac{\theta_i}{2}}\\
        -\sin{\frac{\theta_i}{2}}
        &e^{-i\phi_i}\cos{\frac{\theta_i}{2}}
        \\
        \end{array}
        \right), \label{qstrategy}
\end{equation}
particularly, $\hat{C}= \hat{U}(0,0)$ and $\hat{D}=\hat{U}(\pi,0)$ reduce to the classical strategies. Finally, an operator that destroys the entanglement generated by $\opj(\delta)$ is applied before projecting the output state onto the usual  4-dimensional space basis, giving rise to a probability distribution above the four possible classical states, from which  the expected payoff for each player is determined.
In this paper, we analyse the PD game and demonstrate that {\it local}, and even further, {\it separable} quantum input states suffice to achieve an 
advantage over any classical strategy. This result is in contrast with previous approaches to quantum games that consider entanglement or Bell nonlocality as required resources for achieving a quantum advantage~\cite{meyer,jens,pappa,brunner,Du02}.
Our finding is two-fold: First, we show that neither 
nonlocality nor entanglement can be regarded as the underlying fundamental  properties responsible for the PD quantum advantage: we find purely discord-correlated 
states (zero entanglement) that also achieve such an advantage. Second, we show, by extending the set of Werner-like ({\it W-l}) states $\rho_{W{\text -}l}(p)$, that there exist (non-zero discord) input states for which  the discord does not play any role at reaching this  advantage. We also provide  an optical setup that  implements the locally-powered game strategy, and perform numerical experiments that demonstrate our findings. \\
\begin{figure*}[t]
        \begin{center}
         \includegraphics[scale=0.22]{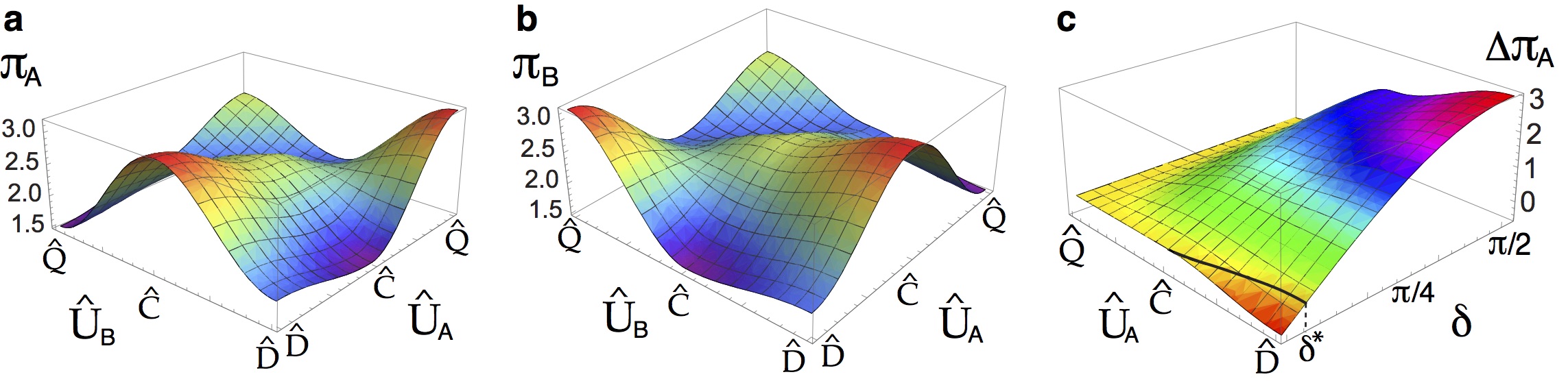}
        \caption{\label{pagos}  {\bf Players payoffs and Nash inequality for the quantum PD game.} ({\bf a})  Alice and  ({\bf b}) Bob's payoff functions for the initial mixed-separable-discorded state 
        $\rho_{in}(p=1/3)$ as function of the strategy space; $(\hat{Q}, \hat{Q})$, with $\hat{Q}=\hat{U}(0,\pi/2)$, is the quantum strategy that removes the dilemma. 
        ({\bf c}) The left-hand-side value in equation~\eqref{nash_ineq}, 
     $\Delta \pi_A$, is plotted as a  function of the players strategies and the measurement parameter $\delta$. The Nash inequality takes positive values almost anywhere the surface, except at the red region below the black curve; e.g, for the particular strategy 
        ($\hat{D},\hat{Q}$), the inequality is not satisfied for $\delta<\delta^*=\arcsin(1/7)$. Since $p$ is just a global factor in  equation~\eqref{nash_ineq}, this behaviour holds even for input states with zero entanglement.}
        \end{center}
\end{figure*}

\noindent
{\bf\large Results}\\
\noindent
{\bf Local quantum correlations as a resource in the PD game.}
In contrast to the use of entangled states as a strategy for `quantising' the PD game~(Fig.~\ref{protocol}(a))~\cite{jens,Du02}, 
we explore a different feature and use the following input states (Fig.~\ref{protocol}(b)) as the feeding resource for performing the quantum PD game:
\begin{align}
\rho_{in}(p,\delta)=\frac{(1-p)}{4}\mathds{I}+p\ket{\psi_{in}(\delta)}\bra{\psi_{in}(\delta)},
\label{wernerstate}
\end{align}
where  $\mathds{I}$ is the $4\times4$ identity matrix, and  $p\in[0,1]$ acts as a  control of the statistical mixture $\rho_{in}(p,\delta)$, and allows a direct comparison with the protocol of 
Fig.~\ref{protocol}(a)~\cite{jens}. In Fig.~\ref{protocol}(b),
the measurement process is made in a basis controlled by the same $\delta$ parameter, which allows the  control of the degree of correlations that is  `destroyed' in the final step of the protocol, just before  the projection onto the usual basis; i.e.,  the quantum operator $\tilde{\mathcal{J}}\equiv\tilde{\mathcal{J}}(\delta)$ inside the dashed rectangle of Fig.~\ref{protocol}(b) is defined in the same way as the entangling operator of  Fig.~\ref{protocol}(a)~\cite{jens}.

Every separable (non-entangled) state is local. However, there exist entangled states which are also local.
For general two-qubit states of the form $\rho:= p \rho' + \frac{(1-p)}{4}\mathds {I}$, $0\leq p\leq1$, being $\rho'$  an arbitrary two-qubit state, a \textcolor{blue}{locality bound} has been reported~\cite{acin06}. 

\begin{figure*}[t]
        \begin{center}
         \includegraphics[scale=0.15]{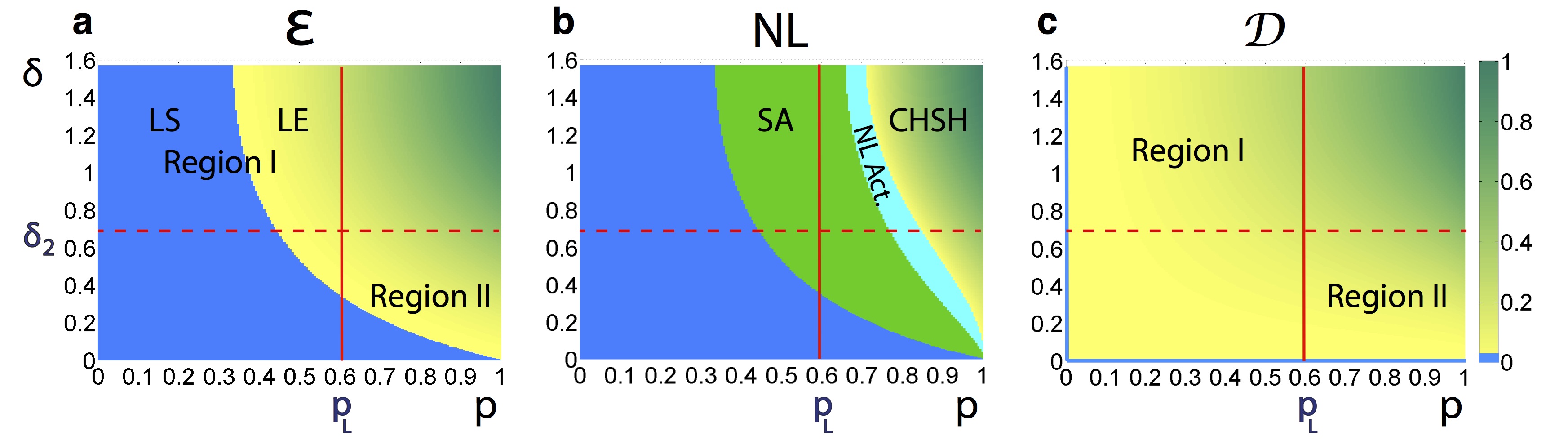}        
        \caption{\label{nonlocal} {\bf Quantum properties of the input states $\rho_{in}(p,\delta)$ and quantum advantage bound.}  As a function of $\delta$ and $p$, we plot:
        ({\bf a}) Entanglement of Formation ($\mathcal{E}$):
        the blue area represents the set of separable and therefore local states, and all the
states $p \leq p_L \approx 0.6009$, as depicted by the vertical line $p = p_L$, are also local (for the joint correlation)~\cite{acin06}; these allow the identification of the local-entangled
(LE) region of states,
        ({\bf b}) non-locality (NL) properties: $\rm CHSH$ inequality violation, $k$-copy nonlocality or superactivation ($\rm SA$) of non-locality (green-solid area), and activation of non-locality (NL Act.) through tensoring and local filtering 
        (cyan-solid area), and  ({\bf c}) quantum discord ($\mathcal{D}$): 
the Region I ($\delta \geq \delta_2$, $p \leq p_L$, upper left rectangles) spans non-zero discord 
        states that even though local, allow a quantum advantage; the Region II 
        ($\delta < \delta_2$, $p > p_L$, lower right rectangles) portrays non-local and local non-zero discord states for which the choice dilemma is not removed.       The bound $\delta \geq \delta_2 = \sin^{-1}\sqrt{2/5}$, for which the quantum advantage holds,  is 
        depicted by a horizontal red-dashed line.} 
        
        \end{center}
\end{figure*}

In  our protocol, we 
identify $\rho'= \ket{\psi_{in}(\delta)}\bra{\psi_{in}(\delta)}$ such that $\rho\equiv \rho_{in}(p,\delta)$, and hence the locality bound reads $p_L\approx 0.6009$; i.e., entangled states with $p \leq p_L$ are local (see  the full local-entangled (LE) region  in Fig.~\ref{nonlocal}(a)). Furthermore, we also account for the set of local,  but separable (LS) states (see the blue Region in Fig.~\ref{nonlocal}(a)).

 \begin{figure*}[]
        \begin{center}
         \includegraphics[scale=0.23]{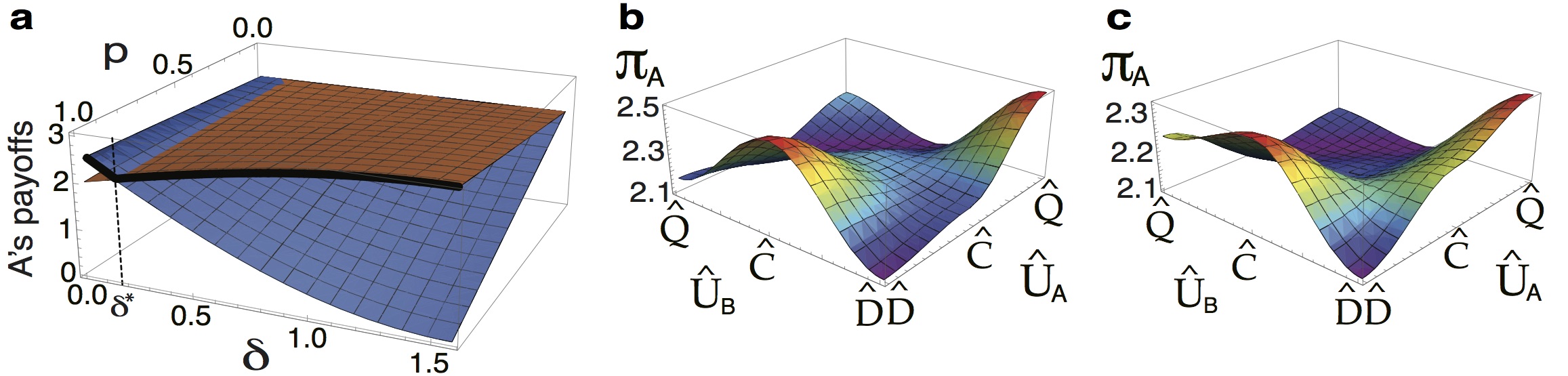}
        \caption{\label{Nash} {\bf Nash equilibrium analysis for the Werner-like initial state:} ({\bf a}) Player $A$'s payoffs for ($\hat{Q},\hat{Q}$) (brown-upper), and ($\hat{D},\hat{Q}$) (blue-lower) strategies as functions of both the entanglement $\delta$ and the mixing  $p$  parameters. The black-solid curve at $p=1$ shows the behaviour of the Nash equilibrium before and after the critical point $\delta^*=\sin^{-1}(1/7)$ (vertical-dashed line). Strategies space profile for player $A$ payoffs with ({\bf b}) $\delta=0.2>\delta^*$, and ({\bf c}) $\delta=0.05<\delta^*$ for the mixed input state $\rho_{in}(p=1/3)$.}
        \end{center}
\end{figure*}

In what follows, we first specialise to {\it W-l} states 
$\rho_{in}(p):=\rho_{in}(p,\pi/2)\equiv\rho_{W{\text -}l}(p)$ as  inputs, and the correlation parameter $\delta$ is fixed to $\pi/2$ for the initial state, and 
 only varied at the measurement. We then generalise our results to input states $\rho_{in}(p,\delta)$, and  consider the 
$\delta$ parameter  being varied at both the input state and the final measurement process. For comparison, we also compute  metrics  to quantify
quantum correlations such as discord $ \mathcal{D}$, entanglement of formation $\mathcal{E}$, and CHSH-nonlocality; see Methods section for definitions.
\\

\noindent
{\bf Quantum local PD  payoffs for the Werner-like states.} 
The  quantum  properties of the states 
 $\rho_{in}(p)$ are shown in Fig.~\ref{protocol}(c), where several distinctive regions can  be identified: {\it local-separable} ($0\leq p \leq p_\mathcal{E}=1/3$), local-entangled ($1/3< p \leq p_L\approx 0.6009$), and entangled-nonlocal ($p\geq p_{NL}=1/\sqrt{2}$) states (see Methods section). Furthermore,  the {\it W-l} states  also highlight quantum correlations at zero entanglement ($p\leq 1/3$)\;\cite{Wer89}, which are captured here  by means of the discord~\cite{Zur01,modi,jh14}. Building on this, we take an approach that is not based on entangled~\cite{jens} or nonlocal~\cite{pappa,brunner} input states. Instead, we choose  local-separable $\rho_{in}(p \leq 1/3$) input states ($ \mathcal{E}=0$, Fig.~\ref{protocol}(c)), while the players' quantum moves  remain ruled by  $\hat{U}_{i}$,  to test whether a quantum strategy based on such states removes the choice problem in the PD game.  
 
We calculate the corresponding PD payoffs for the {\it W-l} input states; for player $A$ this reads:
\begin{widetext}
\begin{eqnarray}
\label{werner1}
\pi_A[\theta_{i},\phi_{i}, p, \delta] &=& \frac{9}{4}-\frac{p}{4} \Big\{f(\theta_{AB})+g(\theta_{AB})\sfp+4\sdp\Big[f^2\Big(\frac{\theta_{AB}}{2}\Big)\cos{2(\phi_A+\phi_B)} - g^2\Big(\frac{\theta_{AB}}{2}\Big) + \\ &&
\frac{5}{2}\Big(\partial_{\frac{\theta_B}{2}}g\Big(\frac{\theta_{AB}}{2}\Big)\Big)^2\cos{2\phi_B}-\frac{5}{2}\Big(\partial_{\frac{\theta_A}{2}}g\Big(\frac{\theta_{AB}}{2}\Big)\Big)^2\cos{2\phi_A} - g(\theta_{AB})\Big(\frac{3}{4} \sin{(\phi_A-\phi_B)}+\partial_{\phi_B}g(\phi_{AB})\Big) \Big]\Big\}, 
\nonumber
\end{eqnarray}
\end{widetext}
with $f(\theta_{AB}):=\ctsp{A}\ctsp{B}$ and $g(\theta_{AB}):=\stsp{A}\stsp{B}$. Player $B$'s payoff is obtained from equation~\eqref{werner1} by exchanging indexes $A$ and $B$ ($i=A,B$).

In Fig.~\ref{pagos}, we plot the  players' payoffs as function of their strategies. We obtain a payoff distribution for which the solution criteria can be evaluated in order to find equilibrium strategies~\cite{Me91}; the classical solution criteria remain valid in the quantum context, and thus we introduce a $(\hat{Q},\hat{Q})$ strategy  that removes the choice problem  in the PD game~\cite{jens}. This result arises `naturally' by fixing $\delta=\pi/2$ and $p=1$ at both the input state and the measurement stage of our protocol.
A thorough examination of the payoff functions, equation~\eqref{werner1}, reveals  that whilst $p$ controls the magnitude of the 
players' payoff, $\delta$ modifies the shape of their distributions. This demonstrates that our {\it local} input state ($p\leq1/3$) keeps unaffected 
the equilibrium properties of the quantum version of the PD game as shown in Fig.~\ref{pagos}(a) and (b) for the particular case $p=1/3$ and 
$\delta=\pi/2$.  We then ask what happens to the Nash Equilibrium if both $p$ and $\delta$ are modified at a given  time,  for which we 
next compute the corresponding Nash inequality.
\\

\noindent
{\bf Nash equilibria of the game.} 

In a finite game of normal form $\left\{ N,\{S_i\}_{i=0}^{n},\{ \pi _i\}_{i=1}^n  \right\}$, 
a strategy chain $s^*$ is NE to the $i$-th player if and only if
$\pi_i(s^*_i,s^*_{-i})\geq\pi_i(s_i, s^*_{-i}), \forall s_i \in S_i$,

where $S_i$ is the strategy space of player $i$, and $\pi$ denotes the payoff function.

We evaluate this criterion with respect to the quantum strategy $(\hat{Q},\hat{Q})$, and for player $A$ we obtain
{\small 
\begin{eqnarray}
\label{nash_ineq}
&&\Delta\pi_{A}:=\pi_{A}(\hat{Q},\hat{Q})-\pi_{A}(s_i,\hat{Q})=\\ 
&& p\Big[\sin{\delta}\Big(1+\cos^2\frac{\theta_A}{2}\cos2\phi_A+\frac{5}{2}\sin^2\frac{\theta_{A}}{2}\Big) +\frac{\cos{\theta_A}-1}{4}\Big]\geq 0. \nonumber
\end{eqnarray}}
Note that $\pi_{A}(\hat{U},\hat{U})\equiv\pi_{A}(\theta_{i},\phi_{i}, p, \delta)$. We reach the same result for $\Delta\pi_{B}$. This inequality does not depend on the value of $p$, and hence it holds even for zero entanglement input states; since we are interested in the quantum case, $p=0$ is discarded.  Thus,  $\delta$ becomes  a crucial factor when  deciding whether  this Nash inequality  is satisfied. 
We highlight the novelty of the result equation~\eqref{nash_ineq}: the quantum advantage, here reported, {\it does not} require neither the maximal entanglement condition $\delta=\pi/2$ (nor any $ \mathcal{E}>0$ at all), nor that of nonlocality to be fulfilled; instead, the quantum strategy $(\hat{Q}, \hat{Q})$ is a NE when the player $A$ moves its strategy from $\hat{C}$ to $\hat{D}$, for  $\sin{\delta}\geq f(\theta)=(1-\cos{\theta})/(11-3\cos{\theta})$. For the specific strategy ($\hat{D},\hat{Q}$), the critical value $\delta^*$ is given by $\sin{\delta^*}= f(\pi) =1/7$, as explicitly shown in Fig.~\ref{pagos}({\bf c}). 
The same result is achieved by analysing the inequality $\Delta\pi_{B}$ obtained for player $B$'s payoff.

For clarity, in Fig.~\ref{Nash}(a) we plot the player $A$'s payoff for the particular ($\hat{Q},\hat{Q}$) (brown-upper) and ($\hat{D},\hat{Q}$) 
(blue-lower) strategies,  in terms of the entanglement $\delta$ and mixing $p$ parameters. The vertical-dashed line on the 
$p=1$ plane marks the critical $\delta^*$ at which the dominant strategy, i.e., the strategy giving a NE, changes. Hence, two regions arise for 
any $p>0$: 
i) $\delta\geq\delta^*=\sin^{-1}(1/7)$, the quantum strategy $(\hat{Q},\hat{Q})$ is the NE and Pareto optimal such that the choice dilemma is 
removed as can be seen for $\pi_A$ in Fig.~\ref{Nash}(b); 
ii) $\delta<\sin^{-1}(1/7)$, the game does not present a strict NE but two at $(\hat{Q},\hat{D})$ and $(\hat{D},\hat{Q})$, the payoff for player $A$ 
is greater when choosing the former rather than the latter strategy, as shown in Fig.~\ref{Nash}(c) (the opposite arises for player 
$B$--not shown). This asymmetry implies again a choice problem in the game such that the dilemma is not removed in this region. 
Figures \ref{Nash}(b) and (c) have been obtained for $p=1/3$, and show that the advantage over any classical strategy is still achieved for separable states.
 \begin{figure}[htb]
        \begin{center}
         \includegraphics[scale=0.10]{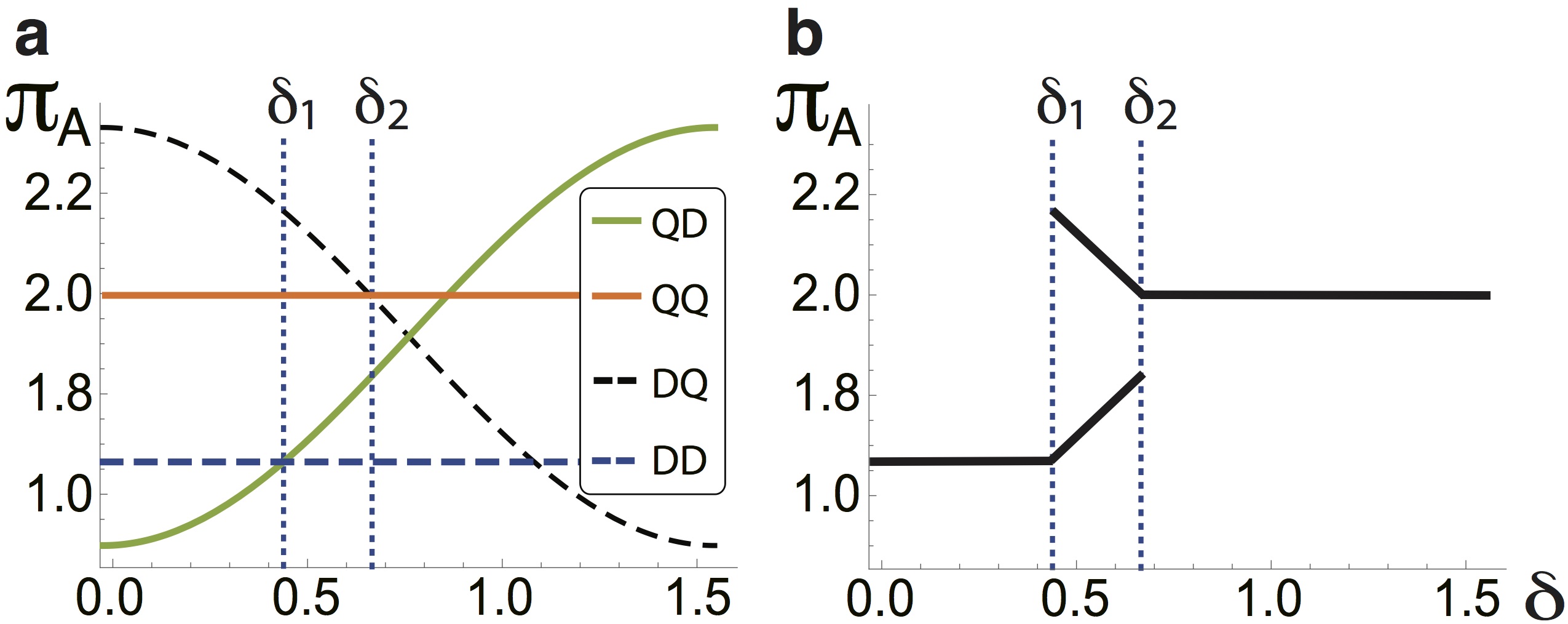}
        \caption{\label{Nash2} {\bf Payoffs for general states $\rho_{in}(p,\delta)$:} ({\bf a}) the control of the initial state correlations,  and $\tilde{\mathcal{J}}(\delta)$ imply  thresholds  at $\delta_1=\sin^{-1}\sqrt{1/5}$, and $\delta_2=\sin^{-1}\sqrt{2/5}$, ({\bf b}) strategies reaching 
        the Nash equilibrium in the regions defined  by $\delta_1$ and $\delta_2$.}
        \end{center}
        \vspace{-0.5cm}
\end{figure}

We stress that the quantum advantage in the PD game, here reported, is not a consequence of entanglement at the input state of the game. In general, as long as $\rho_{in}(p)$ can be generated, the quantum solution for removing the prisoners' dilemma is achieved.  
This means that, for these particular input states, the quantum advantage in the non-zero sum game has been extended to a more general kind of quantum correlations, beyond entanglement,  here quantified by the quantum discord. This is indeed emphasized, as mentioned above, by the quantum properties displayed by the states~equation~\eqref{wernerstate}, as plotted in~Fig.~\ref{protocol}(c) for $\delta=\pi/2$. Indeed, for $p\leq p_L$, $\rho_{in}(p)$ is local; furthermore, if the  resource states  $p\leq 1/3$, then the input states are  local-separable and not related whatsoever to either entanglement or nonlocality. 
In~Fig.~\ref{protocol}(c), we also find that discord is present in the whole $p$-region  $0<p\leq 1$.
This said, a new question arises: how essential is quantum discord as a resource for the quantum advantage here reported?~To address this question, we extend our analysis to input states with a  more general structure, as given by  $\rho_{in}(p,\delta)$. 
\\
 \begin{figure*}[t]
        \begin{center}
         \includegraphics[scale=0.32]{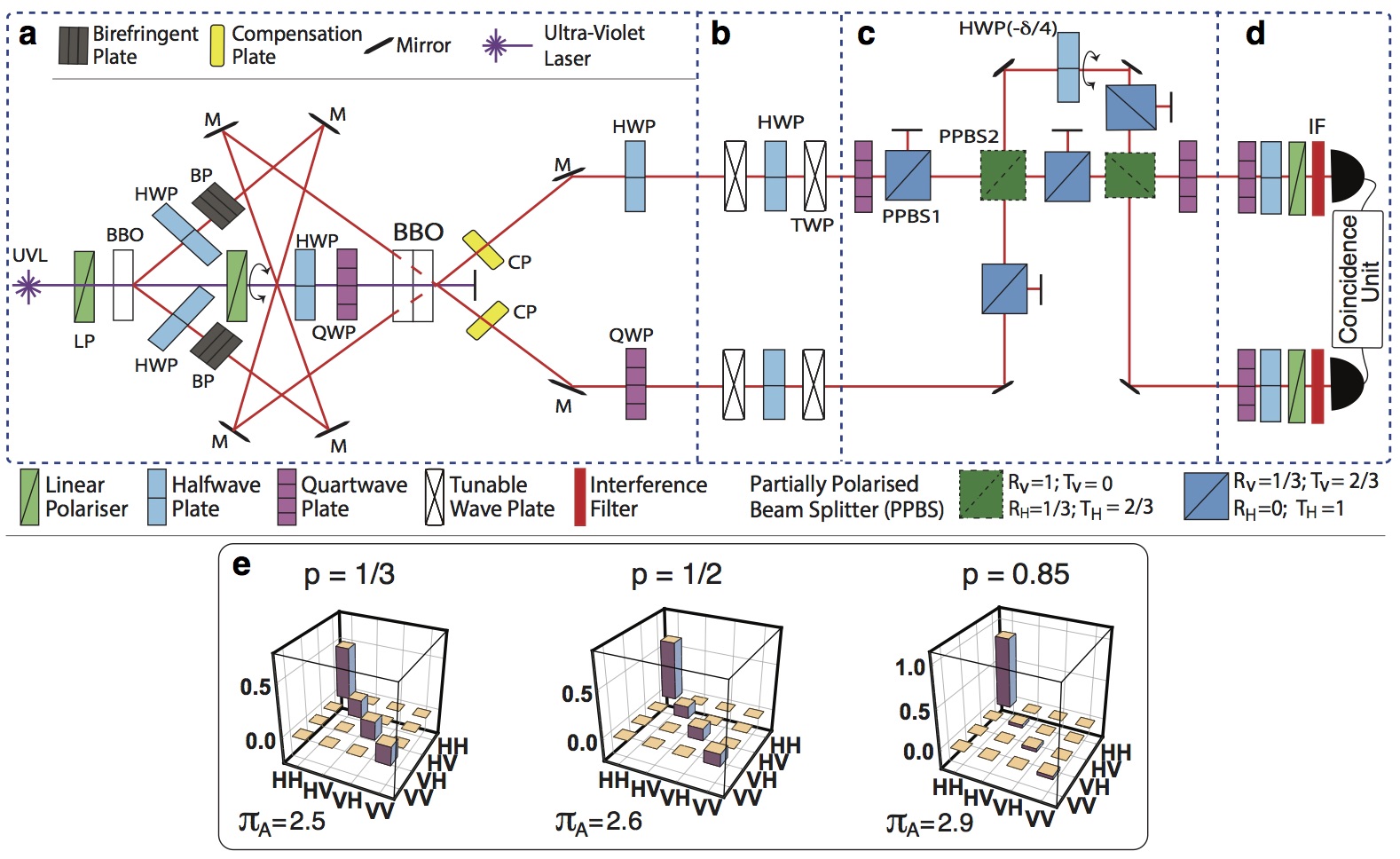}
        \caption{\label{setup}  {\bf Experimental setup to demonstrate the local quantum advantage in the PD game.} Dashed boxes: ({\bf a}) protocol that generates the input states starting from $|VV\rangle$: a Werner state is created 
and successive applications of a $\sigma_x$ and a $\pi$-phase gates lead to $\rho_{in}(p)$ (equation~\eqref{wernerstate}), ({\bf b})  the individual action of  the players on each qubit, $\hat{U}_A$ and $\hat{U}_B$, ({\bf c})  implementation  of the quantum operations $\pi/2$-phase shift, C-NOT, $e^{-i\frac{\delta}{2} Y}\cdot Z$, C-NOT, $\pi/2$-phase shift ($Y$ and $Z$ are the usual Pauli gates), ({\bf d})  the standard tomography protocol to reconstruct the final state which gives the  players payoffs, ({\bf e}) expected tomographies and player $A$'s payoffs for separable ($p=1/3$), local-entangled 
($p=1/2\leq p_L$), and non-local ($p=0.85>p_{NL}=1/\sqrt{2}$) input states; $\delta=\pi/2$, and chosen strategy ($\hat{Q},\hat{Q}$).}
        \end{center}
\end{figure*}

\noindent
{\bf Generalisation to input states $\rho_{in}(p,\delta)$.} If we now control the  input state degree of correlations by varying  $\delta$ in equation~\eqref{wernerstate}, Nash inequality holds as follows: for the strategy $(\hat{D},\hat{Q})$ (or equivalently, for $(\hat{Q},\hat{D})$),
$
 \Delta\pi_{A}:=\frac{p}{2}\left( -3+5\cos2\delta_1\right)\ge0
$,
and  for the strategy $(\hat{Q},\hat{Q})$,
$
 \Delta\pi_{A}:=\frac{p}{2}\left( 1-5\cos2\delta_2\right)\ge0
$.
Three regions arise,  as indicated in Figs. \ref{Nash2}(a) and \ref{Nash2}(b), by means of  $\delta_1=\sin^{-1}\sqrt{1/5}$, and $\delta_2=\sin^{-1}\sqrt{2/5}$. The payoff for the players in the $(\hat{Q},\hat{Q})$ strategy will be constant in the same way that for the $(\hat{D},\hat{D})$ strategy. This behaviour is crucial  for values greater than $\delta_2$ because the Nash equilibrium is reached, and the dilemma is removed. The key parameters  $\delta_1$ and  $\delta_2$ obtained here for the considered mixed states coincide with those reported by Du {\it et al.}~\cite{Du02} for just pure states. This is because $p$ only  affects the size but not the shape of the payoff functions.
For example, by  computing the Nash inequality  for $A$'s payoff   in the $(\hat{Q},\hat{Q})$ and ($\hat{D},\hat{Q}$) strategies, $p$ holds as a global parameter and does not affect the bounds of the inequality. Finally, we show that by considering the {\it W-l} states $\rho_{in}(p)$ and just controlling the degree of correlations  in the final operator  $\tilde{\mathcal{J}}(\delta)$, we reach the quantum advantage which removes the game dilemma for $\delta$ values smaller than those reported before~\cite{Du02}, and, crucially,  $\delta^*<\delta_2$, even for separable states.

For the sake of completeness, we analyse the quantum advantage in the PD game, i.e., the two regions 
defined  by the  $\delta_2$ bound, from which the quantum ($\hat{Q},\hat{Q}$) strategy removes the dilemma, in terms of the quantum properties of 
the input $\rho_{in}(p,\delta)$ states. We plot the entanglement of formation (Fig.~\ref{nonlocal}(a)), 
non-locality given by 
CHSH inequality violation, $k$-copy nonlocality (SA)~\cite{palazuelos},  and activation through tensoring and local filtering~\cite{liang12} 
(NL Act.) (Fig.~\ref{nonlocal}(b)), 
and quantum discord (Fig.~\ref{nonlocal}(c)), all of them as functions of the correlation  $\delta$, and mixing $p$ parameters (see the Methods section for definitions). 
We distinguish two principal regions in Fig.~\ref{nonlocal}: Region I ($\delta \geq \delta_2$, and $p \leq p_L$, upper left rectangles) in 
which it is possible to find local-entangled states, and more interestingly, separable states which are able to remove the choice dilemma 
as they admit the quantum ($\hat{Q},\hat{Q}$) strategy to be the NE and Pareto optimal (see Fig.~\ref{Nash2}). This implies that there exist 
local quantum states that can be seen as a powering resource for performing quantum strategies that outperform any possible classical strategy 
in a PD game. 
In Region II 
($\delta < \delta_2$, and $p > p_L$, lower right rectangles), there are states with different nonlocal properties (Fig.~\ref{nonlocal}(a) and (b)) 
admitting no quantum advantage for removing the choice dilemma in the PD game.
It is worth pointing out that the nonlocal properties here analysed are bounded by entanglement, i.e.,  all of them cover sets 
of states smaller than or equal to the one representing the entangled states.  On the other hand, Fig.~\ref{nonlocal}(c) clearly  shows 
that even for some discord-correlated states, the dilemma is not removed in this region, hence explicitly  showing the existence of non-zero discord states that exhibit no quantum advantage.
Thus, discord on its own cannot be regarded as a fundamental measure (beyond entanglement) that  underpins the quantum advantage. 
\\

\noindent
{\bf Experimental proposal for  demonstrating the locally-correlated quantum advantage.} The described  quantum PD game based on local input states can be experimentally tested, e.g.,  by optical means. In Fig.~\ref{setup} we give

a setup that uses an optical encoding of horizontal ($|H\rangle$) and vertical ($|V\rangle$) polarisation states as qubits. 

The experimental process is divided into four main steps:   preparation of the initial 
state (Fig.~\ref{setup}(a)), setting  the players'  strategies (Fig.~\ref{setup}(b)),  tailoring a quantum operation on the output state 
(Fig.~\ref{setup}(c)), and detection of the game's result via quantum state tomography (Fig.~\ref{setup}(d)). The detailed implementation of these four steps is described in the Methods section.

In Fig.~\ref{setup}(e), we have performed a numerical experiment in order to obtain the Alice's payoffs 
 
based on the local-separable 
 
$\rho_{in}(p=1/3)$, local-entangled $\rho_{in}(p=1/2)$, and the non-local  $\rho_{in}(p=0.85)$ states, for the $(\hat{Q},\hat{Q})$ strategy.
In so doing, we have considered the following feasible experimental parameters:

laser wavelength $\lambda=351$ nm, converted central wavelength $\lambda_0=702$ nm, retardation length $153\, \lambda_0$ and $306\, 
\lambda_0$, spectral bandwidth $\Delta\lambda=10$ nm, and  birefringent plates with a constant difference of $\pi/2$ between them for their 
rotation angles. These simulations are in excellent agreement (not shown) with the result that is obtained by simply following the abstract circuit of Fig.~\ref{protocol}(b). We stress  that our results show that the PD quantum advantage is achieved in the three 
different considered scenarios regardless  the nonlocal or entanglement features of the considered quantum input states.
\\

\noindent
{\bf\large Discussion}
\\
\noindent
Purely {\it local}  and/or {\it separable}  input quantum states have been harnessed as a resource in the PD game, and
we have shown that such a strategy gives a clear advantage over the original bipartite non-zero sum game that makes use of just classical resources. In particular, we have also shown that neither entanglement nor any nonlocal property is strictly required at the input of the game in order to achieve a quantum ($\hat{Q},\hat{Q}$) strategy that removes the PD dilemma and hence outperforms any classical strategy. First, our results have been explored for  Werner-like states with known  nonlocal properties, but also extended to a more general class of correlation-parameter-dependent states (equation~\eqref{wernerstate}). Second, we have shown that within the set of discord-correlated states, there exist some states for which the PD choice problem is not removed, thus implying that quantum discord is neither a necessary condition for achieving the above-described quantum advantage. These results point out the interesting and relevant role played by separable quantum states (and therefore locality) when designing quantum strategies that outperform those based on classical resources, and suggest that such a key resource actually arises from basic quantum interference mechanisms, i.e., quantum coherence, whose description as a physical resource is a rapidly growing conceptual development~\cite{stre}.

The simulated experiment for computing the tomography of the final states of the game, as well as their associated payoff functions (Fig.~\ref{setup}), show that our findings are amenable (although not restricted) to being tested with current photonics technology, as the involved optical devices follow well established, achievable laboratory parameters. We stress that since our PD protocol makes use of disentangled states as captured by equation~\eqref{wernerstate}, their optical generation, via the component 
$\rho'=\ket{\psi_{in}(\delta)}\bra{\psi_{in}(\delta)}$ of the mixed state $\rho_{in}(p,\delta)$, can be facilitated by the fact that `��imperfect'� {\it W-l}  states are more likely to be obtained in the laboratory, in addition to the fact that different $(p,\delta)$-states can be achieved by varying the tilt angle of the second BBO, and by modifying the length of the compensator plates in Fig.~\ref{setup}, thus facilitating the photon interferometry here devised to demonstrate the quantum advantage.

 We remark that we have mainly focused on generating the sufficient conditions for the purely quantum strategy $(\hat{Q},\hat{Q})$ to solve the dilemma in a realistic scenario. 
  This is why we consider an initial state perturbed by a white noise, as well as a non maximally entangled measurement basis. Furthermore, we extend our discussion to the more general case in which not only the entanglement of the measurement basis is varied, but also the entanglement in the $\rho'$  component  of the input state, i.e., we consider the variation of the same correlation parameter $\delta$ at both the beginning and the end of the PD game.
We recently became aware of a report~\cite{Naw} on a related result for the threshold in the NE inequality, but for some particular input entangled states.\\

\noindent
{\bf\large Methods}

  \noindent
{\bf Quantum nonlocality-related properties of the game input  states $\rho_{in}(\delta, p)$.}~A general finite-dimensional bipartite $AB$ system is represented by a density matrix or quantum state $\rho \in D(\mathds{C}^{d_A}\otimes \mathds{C}^{d_B})$, with $d_A, d_B \geq 2$, where $D(\mathds{H}):=\{ \rho \in PSD(\mathds{H})|{\rm{Tr}} (\rho)=1\}$ stands for the set of density matrices of the complex Hilbert space $\mathds{H}$, with $PSD$ the set of {\emph{positive semidefinite}} complex matrices, i.e., the matrices $\rho$ such that $\forall \left| \phi \right> \in \mathds{H}:\left< \phi\right|\rho \left|\phi \right>\geq 0$.  Here, we focus on the quantum properties of our two-qubit input states $\rho_{in}(\delta, p)$ as shown in Fig.~\ref{nonlocal}, where we have emphasised the locality region ($p \leq p_L$) which is limited by the value $p_L \approx 0.6009$ (vertical line), according to the best known bound~\cite{acin06}.   This locality means that a Hidden Variable Model can be found to reproduce the same joint correlation of Alice and Bob $Tr(A\otimes B \rho_{AB})$ predicted by quantum mechanics, where $A$ and $B$ are observables on the state of Alice and Bob, respectively~\cite{acin06}.  The aforementioned nonlocal quantum features
of the input states plotted in Fig.~\ref{nonlocal}  for performing the
PD game are described as follows.\\

 \noindent
{\bf Entanglement.}~We use the entanglement of formation $\mathcal{E}$ as a bipartite entanglement metric~\cite{Woo91}. Let  $\rho_{AB}$ be the quantum state shared by Alice and Bob; the entanglement of formation of $\rho_{AB}$  reads~\cite{Woo91}:   
	\begin{equation}
		\mathcal{E}(\rho_{AB})=h\left(\frac{1}{2}\left[1+ \sqrt{1-\tilde{C}(\rho_{AB})^2}\right]\right),
	\end{equation}
where $h(x)=-x\log_2x-(1-x)\log_2(1-x)$  is the binary entropy, and $\tilde{C}(\rho_{AB})=\max\{ 0,\lambda_4-\lambda_3-\lambda_2-\lambda_1\}$ the \emph{concurrence}. The $\lambda_i$'s refer to the square root of the eigenvalues belonging to the auxiliary operator $\rho_{AB} \tilde{\rho}_{AB}$ arranged in decreasing order, and~$\tilde{\rho}_{AB}=(\sigma_y\otimes\sigma_y)\rho_{AB}^*(\sigma_y~\otimes~\sigma_y)$\;\cite{Woo91}.\\

 \noindent
{\bf Discord.}~The role played by all the quantum correlations in the PD game is cast by means of the quantum discord $\mathcal{D}$, a metric defined as the minimum difference between the quantum version of two classically-equivalent ways of defining the mutual information~\cite{Zur01}:
\begin{eqnarray}
   \label{D}
   \mathcal{D}(\rho_{AB})&=&\text{min}_{\{\Pi_j^B\}}\left(I(\rho_{AB})-J(\rho_{AB})_{\{\Pi_j^B\}}\right) ,
\end{eqnarray}
where $I(\rho_{AB})=S(\rho_A)+S(\rho_B)-S(\rho_{AB})$ is the quantum mutual information, $J(\rho_{AB})_{\{\Pi_j^B\}}=S(\rho_A)-S(\rho_{A|\{\Pi_j^B\}})$ is the conditional mutual information associated to the state of the subsystem (say $A$) after 
the state of the subsystem (say $B$) has been measured (applying POVM operators $\Pi_j^B$), $\rho_{A,B}=\mathrm{tr}_{B,A}(\rho_{AB})$, the conditional entropy $S(\rho_{A|\{\Pi_j^B\}})=\sum_jp_jS(\rho_{A|\Pi_j^B})$, with probability $p_j=\mathrm{tr}(\Pi_j^B\rho_{AB}\Pi_j^B)$, and the density matrix after the measurement on $B$ is given by $\rho_{A|\Pi_j^B}=\Pi_j^B\rho_{AB}\Pi_j^B/\mathrm{tr}(\Pi_j^B\rho_{AB}\Pi_j^B)$\;\cite{Zur01,modi,jh14}. 
\\

  \noindent
{\bf CHSH-Nonlocality.}~Given $\rho \in D(\mathds{C}^2 \otimes \mathds{C}^2)$, the Clauser-Horne-Shimony-Holt (CHSH) inequality~\cite{chsh} considers two dichotomic  observables per party (eigenvalues $\pm 1$), namely ($A_1, A_2, B_1, B_2$), and it takes the form:
\begin{eqnarray}
\left|B_\rho(A_1, A_2, B_1, B_2)\right|&&:=
\\ && \left|E_{11}+E_{12}+E_{21}-E_{22}\right|\leq 2, \nonumber \; \; \; 
\label{CHSH}
\end{eqnarray}
where $E_{ij}:= {\rm{Tr}}\left [\left(A_i\otimes B_j\right)\rho \right]$, $i,j=1,2$. It is said that $\rho$ violates the CHSH inequality if and only if $M(\rho):=\mu+\widetilde \mu>1$,
where $\mu, \widetilde \mu$ are the biggest two eigenvalues of the matrix $U_\rho:=T_{\rho}^TT_{\rho}\in M_{3\times3}(\mathds{R}) $, with $T_{\rho}:=[t_{nm}]\in M_{3\times3}(\mathds{R}) $, with elements $t_{nm}:={\rm{Tr}}[\rho (\sigma_n  \otimes \sigma_m)]$, $\sigma_k$, $k =1,2,3$, the Pauli matrices. This arises from the fact  that $\max B_\rho:=| {\rm{max }}_{A_1, A_2, B_1, B_2}B_\rho|=2\sqrt{M(\rho)}$\;\cite{horo95}. Then, using the Tsirelson's bound~\cite{tsirelson}, $\max B_\rho\leq 2\sqrt{2}$, it follows $0\leq M(\rho)\leq2$, showing nonlocality in the interval $1<M(\rho)\leq 2$. Instead of~$M(\rho)$, we could work with $B(\rho):=\sqrt{ \max\left \{0, M(\rho)-1\right \}}$ given that, for pure states,  the former equals the concurrence: $C(\left| \psi \right>)=B(\left| \psi \right>)$\;\cite{mira}. However, in order to have a direct comparison with $\mathcal{E}$, in Fig.~\ref{nonlocal}(b), we compute nonlocality through the CHSH inequality, by plotting  ${\rm{CHSH}}(\rho):=h([1+\sqrt{1-B(\rho)^2}]/2)$, where $h(x)$ is the binary entropy.\\


 \noindent
{\bf $k$-copy nonlocality (superactivation).}~Given $\rho~\in~D(\mathds{C}^2\otimes \mathds{C}^2)$, if $\rho$ is useful to teleportation then is $k$-copy nonlocal~\cite{cavalcanti}, i.e., $\rho$ admits \emph{superactivation} of nonlocality~\cite{palazuelos}. Usefulness to teleportation can be numerically tested by computing  the Fidelity of Teleportation, which can be written as $\mathcal{F}(\rho)=\frac{2F(\rho)+1}{3}$, where $F$ denotes the Fully Entangled Fraction~\cite{horo99}, which for two qubits reads $F(\rho)=\max\{\eta_{i}, 0\}$,  with 
$\eta_{i}$'s the eigenvalues of the matrix $M=[M_{mn}]$, of elements $M_{mn}= {\rm{Re}} \left (\left<\psi_m \right| \rho \left|\psi_n \right> \right)$, and $\{\left|\psi_n \right>\}$ the so-called magic basis $\left| \psi_{ab}\right>:=i^{(a+b)}(\left| 0,b\right>+(-1)^a\left|1,1\oplus b \right>)/\sqrt{2}$\;\cite{inclusive02}. $\rho$ is useful to teleportation if and only if $\mathcal{F}>2/3$\;\cite{horo99}. In our case, as  shown in 
Fig.~\ref{nonlocal}(b), the set of states that can be super-activated coincides with the whole set of entangled states (although this fact does not hold in general).\\

  \noindent
{\bf Activation of nonlocality through tensoring and local filtering.}~Given $\rho \in D(\mathds{C}^{d_1} \otimes \mathds{C}^{d_2})$ for subsystems $A$ and $B$ with arbitrary dimensions $d_1$ and $d_2$ respectively and, defining $P_{CHSH}$ as the set of states that do not violate the CHSH inequality, even after local filtering, we say that $\rho \in P_{CHSH}$ admits \emph{activation} of nonlocality through tensoring and local filtering \cite{liang12} if there exists a state $\tau_{\rho} \in P_{CHSH}$ such that $\rho \otimes \tau_{\rho} \notin P_{CHSH}$. The latter is equivalent to have $\rm{Tr} \left(\tau_{\rho} \left(\rho^{\rm{T}} \otimes H_{\pi/4} \right)\right)<0$, with $H_{\pi/4}:=\mathds{I}_2\otimes\mathds{I}_2-\frac{1}{\sqrt{2}}(\sigma_x \otimes \sigma_x + \sigma_z \otimes \sigma_z)$, with ${\rm{T}}$ denoting transposition~\cite{liang12}. A theorem~\cite{liang12} establishes the existence of such matrices $\tau_{\rho}$ in the space $ D\big(\bigotimes_{i=1}^2 (\mathds{C}^{d_i} \otimes \mathds{C}^{2})\big)$ for any entangled $\rho$. Although the existence of such a matrix $\tau_\rho$ is already guaranteed, the theorem does not explicitly tell us how to calculate it. We have numerically tested this activation~\cite{liang12} by looking for a state $\tau_{\rho}$ with positive partial transpose with respect to the first subsystem, $\tau_{\rho} ^{T_1}\geq 0$ (say $\mathds{C}^{d_A} \otimes \mathds{C}^{2}$)\;\cite{peres,horo96}, since this  implies $\tau_{\rho} \in P_{\rm CHSH}$\;\cite{masa06}. Thus, we solved the optimisation problem $\sigma(\rho):={\rm{min}}_{\tau_\rho}\rm{Tr} \left(\tau_{\rho} \left(\rho^{\rm{T}} \otimes H_{\pi/4} \right)\right)$ under constrains $\tau_{\rho} \geq 0 \; \wedge \; \tau_{\rho} ^{T_1}\geq 0$\;\cite{liang12}. Even though the considered activation of the nonlocality region covers the whole entangled states~\cite{liang12}, the region for which we are indeed able to find the ancillary matrix required for the activation is represented by the cyan solid area (which covers the CHSH inequality violation region) in Fig.~\ref{nonlocal}(b).\\

  \noindent
{\bf All-optical setup to demonstrate the locally-powered quantum advantage.}~In  Fig.~\ref{setup}(a), a  laser beam is sent, through a linear polariser defining the input, to the first nonlinear crystal (BBO-$\beta$ barium borate type I) as $|H\rangle$. After the first BBO crystal the state holds $|VV\rangle$,   we then use a couple of half-wave plates (HWP) rotated azimuthally $\theta=\pi/8$ to apply a Hadamard gate to each qubit such that  $|V\rangle$ is transformed into $\frac{1}{\sqrt{2}}(|H\rangle-|V\rangle)$, and hence a superposition of all basis states is generated~\cite{1}. Sequentially, a birefringent environment (a set of quartz or BBO plates) is applied to each photon path and tuned to the maximum decoherence, which only affects the off-diagonal elements  of the density matrix~\cite{1}, thus setting  the state $\rho_1=\mathds{I}/4$.
After the first BBO crystal, the non-converted remaining light is transformed into $\frac{1}{\sqrt{2}}(|H\rangle+|V\rangle)$ by a HWP and pre-compensated through a quarter-wave plate (QWP), then directing it to a second set of BBOs which comprises a couple of crystals with mutually-perpendicular optical axes to create a maximally entangled state~\cite{2}. By combining the rays that passed through the first and  second BBOs, the Werner state
$
\rho=p|\Psi^-\rangle\langle\Psi^-|+\frac{1-p}{4}\, \mathds{I}
$
 is produced, where $|\Psi^-\rangle=\frac{1}{\sqrt{2}}(|HV\rangle- |VH\rangle)$ is a  Bell basis state. 
We next apply a  $\sigma_x$-gate to the upper path  through a HWP with $\theta=\pi/4$ as a rotating angle,  thus transforming  $|\Psi^-\rangle$ into the $|\Phi^+\rangle=\frac{1}{\sqrt{2}}(|HH\rangle+ |VV\rangle)$ Bell state. Then, a $\pi/2$-phase shift gate  is applied to the lower path using a QWP with $\theta=0$ and hence producing $\rho_{in}(p,\pi/2)$ (equation~\eqref{wernerstate}), the input state of our quantum PD game.  Here,  $p$ can be tailored by allowing control of the intensity ratio  between the converted light in the first BBO and the converted light in the second BBOs~\cite{1}: $0\leq p\leq 1$ could be tuned by adjusting the  rotation angle of a linear polariser with respect to its optical axis located on the unconverted path just after the first BBO; $p$ can then be measured from the total irradiance ($I_{T}$) after the second conversion, and the partial irradiance ($I_P$) of the light converted in the first BBO, as $p=\frac{I_T-I_P}{I_T}$.  Thus, the local $\rho(p=1/3)$ input state can be  achieved by setting $I_P=2 I_T/3$.

Figure~\ref{setup}(b) implements the actions of the players (operator $\hat{U}_i$ in  equation~\eqref{qstrategy})  by means of a set of wave plates, where the phase $\phi$ corresponds to the  retarding angle of each plate, i.e., $\phi=\pi$ for a HWP, and $\phi=\pi/2$ for a QWP. The  angle  $\theta$ corresponds to the perpendicular rotation of the centre half wave plates,  referred to their optical axis. In Fig.~\ref{setup}, we  use a special kind of wave plate that does not have a defined angle $\phi$, the so-called  tunable wave plate (TWP), which allows us to generate $0\leq \phi\leq 2\pi$.
In Fig.~\ref{setup}(c) we start from  the output  state $\rho_{out}$ (Fig.~\ref{protocol}(b)) generated in the previous step. To test the NE inequality, equation~\eqref{nash_ineq}, we apply a phase gate by means of a QWP on the lowermost path, and resort to the use of a quantum Controlled-NOT gate 
 
which comprises a set of three partially polarised beam splitters (PPBS)~\cite{Langf05,7,9,10}, where the two PPBS1s completely transmit the photons with $|H\rangle$ and $1/3$ of the $|V\rangle$ polarisation, while the PPBS2 completely  reflects $|V\rangle$ and $1/3$ of the $|H\rangle$ polarisation. 
Then, a HWP with $\theta=-\frac{\delta}{4}$  acts as a controller of the $\delta$ parameter over the control output of the first C-NOT gate (uppermost  path), and additionally, a symmetrical arrangement of another C-NOT,  and a QWP($\theta=0$) completes the quantum operator $\tilde{\mathcal{J}}(\delta)$.
Finally,  the measurement process is depicted in Fig.~\ref{setup}(d);  a standard quantum state tomography protocol~\cite{5}, which requires a set of 16 measures is performed in order to obtain the final state of the system and the result of the game.


\noindent
{\bf\large Acknowledgements}

  \noindent
C.E.S. gratefully  acknowledges V. Vedral and his Group for  hospitality and discussions during a research stay where part of this work was performed. C.A.M.-L. acknowledges to J. K\"ohler, R. Hildner and the EPIV research chair for the valuable support during a research stay.  We are grateful to A. Arg\"uelles for a critical reading of the manuscript.
We acknowledge financial support from the Colombian Science, Technology
and Innovation Fund-General Royalties System (Fondo CTeI-Sistema General de Regal\'ias, contract~BPIN 2013000100007),  COLCIENCIAS (grant~71003), Universidad del Valle (grant~7930) and CIBioFi.
\\

\noindent
{\bf\large Author contributions}

  \noindent
C.A.M.-L. and C.E.S. contributed equally to this work. C.A.M.-L., C.E.S., and A.B. computed the payoffs and Nash equilibrium inequalities. A.F.D. calculated the quantum properties of the considered states. C.A.M.-L. and C.E.S.  developed the experimental proposal. J.H.R. originated the concept, guided  the simulations and performed theoretical analysis. 
All the authors contributed to the writing of the paper, with a major input from C.E.S. and J.H.R.

\end{document}